\documentclass[12pt,preprint]{aastex}
 
\usepackage{emulateapj5}
\usepackage{epsf}
\usepackage{graphics}

\shorttitle{Size evolution of most massive galaxies at $1.7<z<3$}
\shortauthors{F. Buitrago et al.}

\begin{document}      

\title{Size evolution of the most massive galaxies at $1.7<z<3$ from
GOODS NICMOS survey imaging}

\author{Fernando Buitrago$^{1}$, Ignacio Trujillo$^{2}$, Christopher J.
Conselice$^{1}$, Rychard J. Bouwens$^{3}$, Mark Dickinson$^{4}$, Haojing Yan$^{5}$}

\altaffiltext{1}{School of Physics and Astronomy, University of Nottingham, NG7
2RD, UK. Contact address: ppxfb@nottingham.ac.uk}

\altaffiltext{2}{Instituto de Astrof\'{i}sica de Canarias, V\'{i}a L\'{a}ctea
$s\backslash n$, 38200 La Laguna, Tenerife, Spain}

\altaffiltext{3}{Astronomy Department, University of California, Santa Cruz,
CA 95064, USA}

\altaffiltext{4}{National Optical Astronomy Observatory, Tuscson, AZ 85719, USA}

\altaffiltext{5}{The Observatories of the Carnegie Institution of Washington, 813
Santa Barbara Street, Pasadena, CA 91101, USA}

\begin{abstract}

We measure the sizes of 82 massive ($M\geq10^{11} M_{\odot}$)
galaxies at $1.7\leq z\leq 3$ utilizing deep HST NICMOS data taken in the
GOODS North and South fields. Our sample is almost an order 
of magnitude larger than  previous studies at these redshifts, providing
the first statistical study of massive galaxy sizes at $z > 2$, confirming the 
extreme compactness of these galaxies.  We split our sample into disk-like ($n\leq2$) and 
spheroid-like ($n>2$)
galaxies based on their S\'{e}rsic indices, and find that at a
given stellar mass disk-like galaxies at $z\sim2.3$ are a factor of 2.6$\pm$0.3
smaller than present day equal mass systems, and spheroid--like galaxies at the
same redshifts are 4.3$\pm$0.7 smaller than comparatively massive elliptical galaxies
today.  At $z>2$ our results are compatible with both a leveling off, or a mild evolution in 
size. Furthermore, the high density ($\sim$2$\times$10$^{10}$M$_{\odot}$kpc$^{-3}$)
of massive galaxies at these redshifts, which are similar to present day globular clusters, 
possibly makes any further evolution in sizes beyond z=3 unlikely.

\end{abstract}


\section{Introduction}

One of the most exciting discoveries in extragalactic astronomy in the last few
years is that massive ($M\geq10^{11} M_{\odot}$) galaxies at $z>1$ were
extremely compact (Daddi et al. 2005; Trujillo et al. 2006b, 2007; Longhetti et
al. 2007), particularly those with the lowest estimated star formation rates
(Zirm et al. 2007; Toft et al. 2007; Cimatti et al. 2008; van Dokkum et al.
2008; P\'{e}rez-Gonz\'{a}lez et al. 2008). Since only a very few dense and massive objects are found at
z$\sim$0 (Bernardi et al. 2006; cf. with none at r$_{e} < $1 kpc) it is
clear that significant growth in the sizes of these galaxies has occurred
during cosmic history.

Within the current galaxy formation paradigm, the origin of these galaxies can
be described by the collapse and merging of dark matter
haloes. Models suggest that at very early times galaxies contain 
large amounts of cold
gas, resulting in efficient starbursts (e.g., Khochfar \& Silk 2006).
As star formation occurs, the gas in these galaxies becomes heated due to 
various feedback processes (e.g., Granato et al. 2004, Menci et al 2006), leading
to reduced
star formation rates, creating compact and massive remnants. Observationally,
we know
that at $z < 2$ there are few gas rich mergers in massive galaxies 
based on structural analyses (e.g. Conselice et
al. 2003; Conselice 2006; Conselice et al. 2008) preventing significant 
starbursts and new star populations from forming. Since at these
lower redshifts the amount of available gas has decreased, ``dry'' mergers
are expected to be the dominant mechanism for size and stellar mass
growth (Ciotti \& van Albada 2001, Dominguez-Tenreiro et al. 2006;
Boylan-Kolchin et al 2006), although other processes are possible 
(Naab et al. 2007; Pipino \& Matteucci 2008).

One of the ways to trace this evolution is through measuring the sizes of
galaxies through time.  At $z < 2$, the size evolution of the most massive galaxies has 
been well characterized with large samples of objects. Recently, 
Trujillo et al. (2007) using
$\sim$800 sources found that, at a given stellar mass, disk--like objects at
z$\sim$1.5 were a factor of two smaller than their present-day counterparts.
For spheroid--like objects the evolution is even stronger. These spheroidal
objects are a factor of four smaller at $z \sim 1.5$ compared with similar mass modern
ellipticals. This evolution is also in qualitative agreement with  
hierarchical semi-analytical model predictions which find a factor of $1.5-3$ 
evolution in size since that redshift (e.g., Khochfar \& Silk 2006) .

At $z>$2, however, our knowledge of the size evolution of the most massive
objects is much more scarce. There are only a few attempts to explore this
issue  using small
samples of massive galaxies at z$\sim$2.5 (Zirm et al. 2007; Toft et al. 2007; van Dokkum et al. 2008). 
This is due to the intrinsic scarcity of distant massive galaxies, 
and the relative small sizes of 
previous deep NIR imaging surveys. With the aim of substantially increase our knowledge of the
size evolution of massive galaxies in the redshift interval 
$1.7\leq z\leq 3$, we have imaged a sample of 82 very massive galaxies in 
the GOODS North and South
fields  within the H-band filter as part of the GOODS NICMOS Survey 
(Conselice et al. 2008, in prep).

To allow a comparison with both the local SDSS stellar mass-size relations, and
the results obtained at lower redshifts ($z<2$), we split our sample according 
to light
concentration using the S\'{e}rsic index $n$ to separate disk-like galaxies 
from more concentrated spheroid-like systems. We
find that both types continue, and perhaps level off, in their size evolution
at $z > 2$.   We
assume the following cosmology throughout: H$_{0}$=70 km s$^{-1}$Mpc$^{-1}$,
$\Omega_{\lambda}$ = 0.7, and $\Omega_{\rm m}$ = 0.3, and use AB magnitude
units

\section{Data and Sample}

Our sample of galaxies originates from the GOODS North and South fields and
are imaged as part of the  GOODS NICMOS Survey (GNS; PI C. Conselice). 
The GNS is a large HST NICMOS-3 camera program of 60
pointings  centered around massive galaxies at $z = 1.7-3$ at 3 orbits depth,
for a total of 180 orbits in the F160W (H) band. Each tile
(52"x52", 0.203"/pix) was observed in six exposures that were combined to
produce images with a pixel scale of 0.1", and a Point Spread Function (PSF) 
of $\sim$0.3"
Full Width Half Maximum (FWHM). The details of the data reduction procedure are discussed
in Magee, Bouwens \& Illingworth (2007).  We optimize our pointings to obtain as many
high-mass M$_{*}$ $> 10^{11}$ M$_{\odot}$ galaxies as possible, with the selection 
of these targets described in Conselice et al. (2008).  These
galaxies consist of Distant Red Galaxies from Papovich et al. (2006),
IEROs from Yan et al. (2004), and BzK galaxies from Daddi et al. (2007).
Within our NICMOS fields we find a total of 82 galaxies with masses larger than
$10^{11}h_{70}^{-2}M_{\odot}$  with photometric and spectroscopic redshifts in the 
range $1.7\leq z\leq3$. In addition to these data, and to allow a 
comparison with the sizes obtained in the H-band, we measure, whenever 
possible, the 
sizes of the same galaxies using the $z-$band (F850LP, 5 orbits/image) HST ACS 
data. The $z-$band data is drizzled to a scale 0.03"/pix and
has a PSF FWHM of $\sim$0.1$\arcsec$. Limiting magnitudes reached are $H\sim26.8 
(5\sigma)$ and $z=27$ (10$\sigma$ in a 0.2\arcsec aperture) (Giavalisco et al. 2004).

\section{Determination of stellar masses and photometric redshifts}

The masses and photometric redshifts of our objects are calculated using the 
large suite of GOODS data from the B-band to the infrared (e.g., Giavalisco 
et al. 2004). For our work we used the filters BVRIizJHK.
Stellar masses are measured using 
standard multi-color stellar population fitting techniques, producing 
uncertainties of $\approx0.2$ dex.
Details of the procedure for stellar mass determinations are in e.g., Papovich
et al. (2006), Bundy et al. (2006), Yan et al. (2004) and
Conselice et al. (2007 and 2008 in preparation).  Our stellar masses are 
calculated by assuming a 
Chabrier (2003) Initial Mass Function (IMF) and producing model 
Spectral Energy Distributions (SEDs) constructed from Bruzual \& 
Charlot (2003) stellar populations synthesis models parameterized by an 
exponentially declining star formation history. These
model SEDs are fit to the observed SEDs of each galaxy to obtain a stellar
mass.   Issues concerning newer models utilizing AGB stars (see Maraston et al. 2006) are discussed 
in Conselice et al. (2007) and Trujillo et al. (2007), although
we find that these newer models do not significantly alter our measured stellar
masses.

Another source of uncertainty are the photometric redshifts we use
in our sample which originated from standard techniques (e.g., Conselice
et al. 2007). From the literature we find seven spectroscopic
redshifts for our sample.  Using the GOODS/VIMOS DR1 (details in Popesso et al. 
2008) we find three matches with $\delta z/(1+z)=$0.026, 
and four more from the compilation of GOODS-S 
spectroscopic redshifts Wuyts et. al (2008) giving 
$\delta z/(1+z)=$0.034.

\begin{figure*}
\begin{center} 
\vspace{-1cm}
\hspace{-1cm}
\rotatebox{0}{
\includegraphics[width=0.8\linewidth]{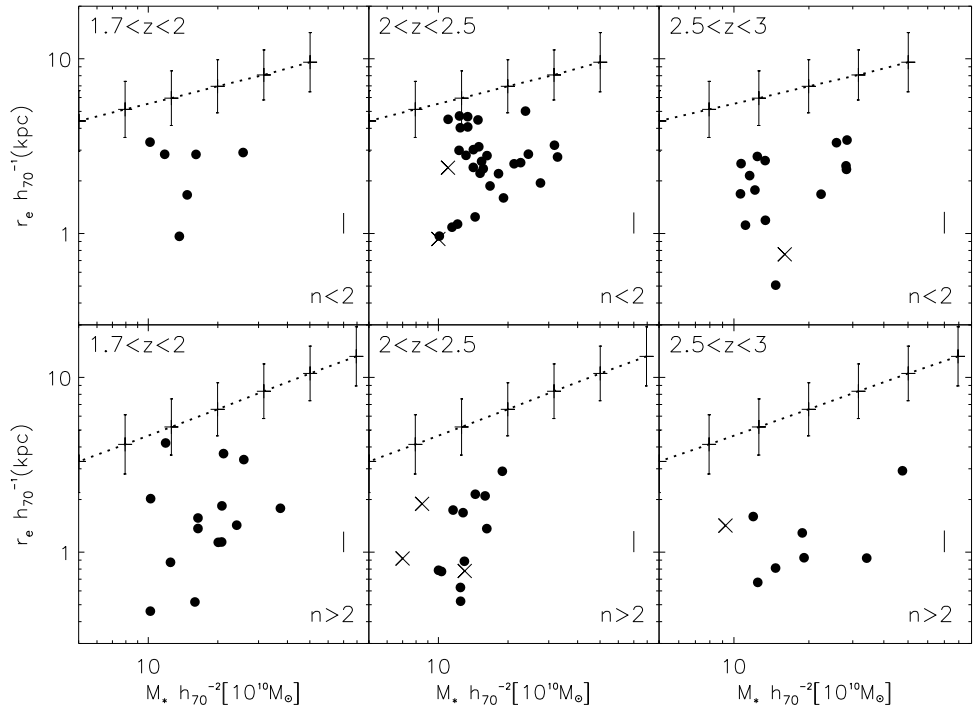}}
\end{center}
\vspace{-0.5cm}

\figcaption{Stellar mass-size distribution for our sample. Overplotted are the
mean and the $1 \sigma$ dispersion of the distribution of the S\'{e}rsic half-light radius
of SDSS galaxies as a function of stellar mass (Shen et al. 2003) and the crosses are the
galaxies from van Dokkum et al. (2008) whose masses have been converted to our IMF.
For clarity, individual error bars are not shown for our data, but a typical size error
bar is shown in the right side of each bin. This mean
size relative error is 0.04" which is 0.32 kpc at z=2.5. Uncertainties in the
stellar masses are $\sim0.2$ dex.}

\end{figure*}

\section{Determination of galaxy sizes}

S\'ersic indices and sizes, as parameterized by the effective radius
along the semi-major axis a$_e$, were measured using the GALFIT code (Peng et
al. 2002). Our measured sizes are circularized, $r_{e}=a_{e}\sqrt{1-\epsilon}$, with
$\epsilon$ the projected ellipticity of the galaxy. GALFIT convolves
S\'ersic (1968) r$^{1/n}$ 2D models with the PSF of the images and determines
the best fit by comparing the convolved model with the observed galaxy surface
brightness distribution using a Levenberg-Marquardt algorithm to minimise the
$\chi^2$ of the fit. We use single S\'ersic models to compare our
size estimations with previous work at lower redshifts. We first estimate 
the apparent magnitudes and sizes of our galaxies using SExtractor
(Bertin \& Arnouts 1996) which were then used as inputs to the GALFIT code.

Before we carry out our fitting we remove neighbouring galaxies 
using an object mask. In the case of very close galaxies with overlapping 
isophotes, objects are fit simultaneously. Due to the point-to-point variation of the shape of
the NIC3 PSF in our images we select five (non-saturated) bright stars to
gauge the accuracy of our parameter estimations. The 
structural parameters of each individual galaxy are measured five times, using
each time a unique star. The uncertainty (1$\sigma$) on the structural parameters due
to changes in the PSF is $\sim$15\% for r$_e$, and $\sim$20\% for the 
S\'ersic index $n$.

Surface brightness dimming is one of the main concerns when measuring
sizes and Sersic indices at high redshift, which in principle could
bias our measured sizes. In previous papers we conduct many
simulations in order to check the importance of surface brightness
dimming at different observational conditions (NIR ground-based;
Trujillo et al. 2004; 2006a,b; and using ACS data Trujillo et al.
2007).   Trujillo et al. (2006a) show through extensive simulations of
galaxies with various sizes and magnitudes, within observing
conditions and depth worse than the NIC3 data we use, that sizes can
be retrieved easily within the magnitude ranges of our objects
($K_{\rm AB} \sim 21.5$).

We check in addition the accuracy of our structural parameter determinations by
comparing our H-band measurements (giving optical rest-frame) against the
results obtained using the $z-$band (NUV rest-frame) from the ACS imaging,
where the spatial resolution is a factor of three better. Unfortunately, 
at $z\gtrsim2$ a large fraction 
(49 out 82) of our galaxies are not detected in the $z-$band, 
and cannot be used for the comparison. For the 33 objects remaining we 
find a good correlation (Pearson correlation
coefficient 0.59) between the sizes measured in both bands, with a small 
possible bias towards smaller sizes (4$\pm$6\%) in the H-band compared to 
the $z-$band measurements. This potential bias towards smaller sizes at longer wavelengths is as 
expected (see e.g. Barden et al. 2005;
McIntosh et al. 2005; Taylor-Mager et al. 2007; Trujillo et al. 2007). Comparing the S\'ersic 
index $n$ is less straightforward, since the patchy distribution of UV light makes the
measurement of the index $n$ (i.e. the shape of the surface brightness profile) very different
from the light coming from more evolved stellar populations. We however find a correlation
between the S\'ersic index as measured in ACS and in NICMOS imaging
(Pearson correlation coefficient
0.36). The S\'ersic indices measured with NICMOS are 13$\pm$12\% smaller 
than those in the ACS $z-$band imaging. Part of the reason for the smaller 
value of the 
index $n$ in the NICMOS images is due to the larger PSF size in the infrared images compared to the PSF in the ACS data.

\section{The observed stellar mass vs size relation}

The stellar mass-size relation for our sample is shown in Figure 1, where we
have split our sample into 3 redshift bins. Overplotted on 
each panel is the local
value of the mean half-light radii and its dispersion at a given stellar mass (based
on Sloan Digital Sky Survey data; Shen et al. 2003). SDSS sizes were 
determined using r'-band data, which is equivalent to the V-band rest-frame at
$z\sim0.1$, the mean redshift of the galaxies in SDSS,  and using a circularized
S\'{e}rsic model. Stellar mass determinations of the galaxies in the local
reference relation were measured using a Kroupa (2001) IMF which gives nearly the
same stellar masses as using a Chabrier IMF.

Our galaxies were split into two types using our measured S\'ersic indices. 
As shown by e.g., Ravindranath
et al. (2004) there is a correlation between the S\'ersic index $n$ and 
Hubble type. Following this correlation, galaxies are usually segregated into
late--type galaxies (with n$<$2-2.5) and early--type (with n$>$2-2.5). The
effect of using either $n=2$ or $n=2.5$ does not significantly alter the 
derived mass-size relation for the local galaxies. We use $n = 2$ as our limit
to account for the
systematic bias towards smaller values of the measured S\'ersic index  
when the PSF size is similar to the sizes of the objects measured (see a
detailed explanation of this effect in Trujillo et al. 2006a; see also Marleau
\& Simard 1998). To obtain a realistic comparison between the local
relation and the size evolution of massive galaxies at $z < 2$, Trujillo et al. (2007) 
use $n$=2.5 to separate disk-like and spheroid-like systems. Trujillo et al. (2007)
use a larger S\'{e}rsic index for their cut, as their PSF sizes are much smaller than 
the galaxies they measure,  unlike in our present sample. Our
choose of $n=2$ is reinforced by exploring the S\'ersic distribution of the index
$n$ in our sample, where two peaks are found at n$\simeq$1 and at n$\simeq$2.3-2.5.

Figure 1 shows that at a given stellar mass our massive galaxies are 
progressively smaller
at high--z. Remarkably, none of our galaxies at $z>1.7$ fall in the mean
distribution of the local relation. Moreover, if the stellar masses were overestimated by a factor of 
two, only three galaxies from our sample would fall in the dispersion of the 
local relation, showing the reliability of our 
results in spite of the uncertainties.  To quantify  the observed size evolution,
we calculate the ratio between the sizes we measure, and the measured sizes
of nearby galaxies at the same mass, by using the SDSS results (Shen et al.
2003). We perform a linear interpolation between SDSS points when necessary.
The evolution of the median ratio is shown on Figure 2, and listed in Table
1. Each point represents the median, and the errors bars the uncertainty, 
on this value ($1 \sigma$). We
also plot the  SDSS reference point and the values obtained by Trujillo et al.
(2007) in the redshift range 0.2$<$z$<$2.   We fit the evolution of the decrease
in half-light ratio with redshift as a power-law $\sim \alpha(1+z)^{\beta}$, where
we calculate that for the disk-like galaxies $\beta = -0.82\pm0.03$ and for the 
spheroid-like systems $\beta = -1.48\pm0.04 $ (Figure~2). This shows that the 
spheroid-like galaxies have a faster rate of decline in size than the disk-like systems.

\begin{figure*}
\begin{center}
\vspace{0cm}
\hspace{-1cm}
\rotatebox{0}{
\includegraphics[width=0.6\linewidth]{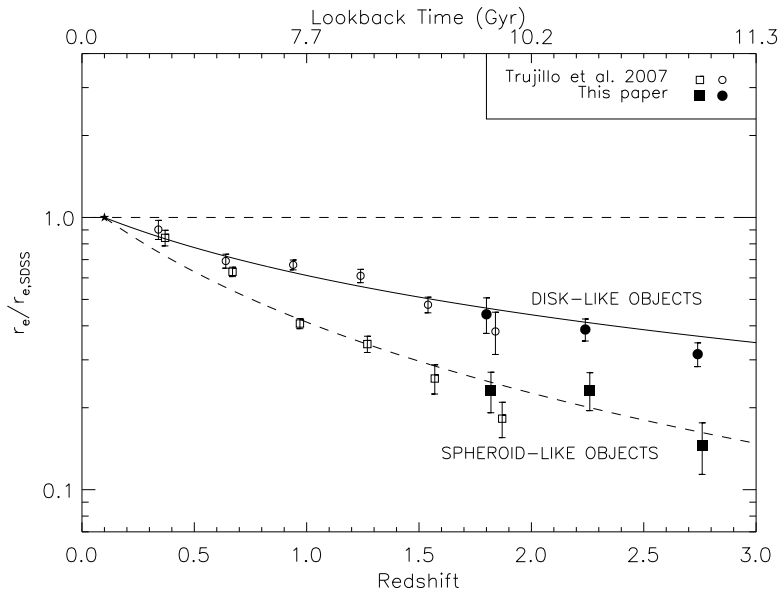}}
\end{center}
\vspace{0cm}

\figcaption{Size evolution of massive galaxies (M$_{*}$ $> 10^{11}$ M$_{\odot}$) 
with redshift. Plotted is the ratio of the median sizes of galaxies in our sample 
with respect to sizes of nearby galaxies in the SDSS local comparison (solid points). 
The results of Trujillo et al. (2007) for systems at 0.2$<$z$<$2 are overplotted (open squares).  
The error bars indicate the uncertainty (1$\sigma$) at the median position.}

\end{figure*}   


Furthermore, as we can see in Figure 2 our results are in agreement with previous work from
Trujillo et al. (2007), and at all redshifts the spheroid-like objects are on average 
smaller than the disk-like galaxies. As Figure~2 shows, we find that disk--like galaxies 
and spheroid--like galaxy decrease only slightly in size beyond z=2.  
Although we see some evolution at $1.7 < z < 3$, the significance of this
is 2.2$\sigma$ for disks and  $1.8 \sigma$ for spheroids (Table~2), 
implying that a flat evolution is possible.
The internal stellar mass densities of the spheroid--like objects 
in our sample at z$>$2 are more
than two orders of magnitude larger than objects of the same stellar mass today (similar to the
results found by van Dokkum et al. 2008).

\section{Discussion}

As has been demonstrated in previous work, some massive galaxies at $z < 2$ grow 
in size by up to
an order of magnitude (e.g., Trujillo et al. 2007).  An interesting question  
is whether these massive galaxies become progressive smaller at higher
redshifts, containing possibly even smaller sizes at $z>2$. 
Our results provide the first statistical sample in which to answer
this question.  As seen in Fig. 2, the objects in our sample
are compatible with the idea that the size evolution reaches a plateau
beyond $z=2$. To shed some light on this question we compute the stellar mass density of
our galaxies and compare these to the densest collection of stars in the local
Universe -- globular clusters. This comparison is an interesting one, since globular clusters
are also expected to be form either very early, or more recently as a result of mergers
of gas clouds during galaxy collisions.


A typical spheroid--like galaxy in our sample at z$\sim$2.75 has a stellar mass of
$\sim$2$\times$10$^{11}$M$_{\odot}$, and a size of r$_e$$\sim$1 kpc. The stellar density for this object,
assuming spherical symmetry, is $\rho=(0.5M)/(4/3\pi
r_{e}^{3})$$\sim$2.4$\times$10$^{10}$M$_{\odot}{\rm kpc}^{-3}$. A disk--like galaxy 
at z$\sim$2.75 has a typical mass of $\sim$2$\times$10$^{11}$M$_{\odot}$ and size r$_e$$\sim$2 kpc. Assuming a disk symmetry, the stellar mass density within 
these disk-like systems is
$\rho=(0.5M)/(\pi r_{e}^{2}h)$$\sim$2.6$\times$10$^{10}$M$_{\odot}{\rm kpc}^{-3}$, where we have used h$\sim$0.3 kpc.
In both cases the stellar mass densities are similar. A typical globular cluster (r$_e$=10
pc and M$\simeq$10$^5$M$_{\odot}$) has a density of
$\sim$1.2$\times$10$^{10}$M$_{\odot}$kpc$^{-3}$. This is remarkably similar to our massive
galaxies at $z > 2$, and reveals that these high--z galaxies may in principle have an origin 
similar to globular clusters.
These high densities also suggest that their stellar mass densities
likely do not become much larger at high redshifts ($z > 3$).  
A massive galaxy at $z > 2$ must also have
formed very quickly, and consequently these high stellar densities could 
reflect the high gas densities in the primeval Universe. The
compactness of our objects, and their similar densities to globulars, is consistent with a scenario 
whereby more massive haloes start collapsing earlier and drag along a large amount of 
baryonic matter that later forms into stars.


If, as suggested by the high density of our galaxies, the size evolution is stopped or
diminished at $z>2$, this perhaps reveals a different evolutionary mechanism for massive
galaxies at $z<2$. A faster size evolution
is in agreement with theoretical models, which predict that the amount of gas involved
in galaxy mergers decreases with lower redshifts. A lower amount of gas results in a
more efficient size growth, as the energy of the collision is not dissipated into the formation of
new stars (e.g., Khochfar \& Silk 2006).

We thank S\'{e}bastien Foucaud and Asa F. L. Bluck for instructive discussions, and
Leonel Guti\'{e}rrez and In\'{e}s Flores Cacho for their valuable help with 
computing issues.  We also thank the other members of the GNS team, 
particularly Emanuele Daddi and Casey Papovich for their 
participation in the various aspects of the survey. The GNS is supported by
NASA grant HST-GO-11082.

\renewcommand{\arraystretch}{1.2}
\begin{center}
\begin{table*}

  \caption{Size evolution of massive (M$_{\star} > 10^{11}$
M$_{\sun}$) galaxies at 1.7$<$z$<$3}
\hspace{1.6in}
\begin{tabular}{ccc}
  \hline
Redshift Range & n$<$2 &  n$>$2   \\
               &  $<$r$_e$/r$_{e,SDSS}$$>$($\pm$1$\sigma$) 
					 &  $<$r$_e$/r$_{e,SDSS}$$>$($\pm$1$\sigma$) \\
 \hline
0.1 (SDSS)& 1 		& 1           \\
1.7-2.0	  & 0.44(0.07)  & 0.23(0.04)  \\
2.0-2.5	  & 0.39(0.04)  & 0.23(0.04)  \\
2.5-3.0	  & 0.31(0.03)  & 0.14(0.03)  \\

\hline
\label{data1}
\end{tabular}
\end{table*}
\end{center}
\renewcommand{\arraystretch}{1}

\renewcommand{\arraystretch}{1.2}
\begin{center}
\begin{table*}

  \caption{Fit to $\sim \alpha(1+z)^{\beta}$ to Trujillo et al. (2007), our data and Figure 2}
\hspace{2in}
\begin{tabular}{ccc}
  \hline
Disk--like galaxies \\
 \hline
 \hline
Redshift & $\alpha$($\pm$1$\sigma$) &  $\beta$($\pm$1$\sigma$)\\
 \hline
0.0-2.0	  & 1.08(0.02)  & -0.78(0.04)  \\
1.7-3.0	  & 1.85(0.28)  & -1.34(0.59)  \\
0.0-3.0	  & 1.08(0.01)  & -0.82(0.03)  \\

\hline
  \hline
Spheroid--like galaxies\\
 \hline
 \hline
Redshift & $\alpha$($\pm$1$\sigma$) &  $\beta$($\pm$1$\sigma$)\\
 \hline
0.0-2.0	  & 1.16(0.01)  & -1.51(0.04)  \\
1.7-3.0	  & 1.42(0.29)  & -1.66(0.92)  \\
0.0-3.0	  & 1.15(0.01)  & -1.48(0.04)  \\

\hline
\label{data2}
\end{tabular}
\end{table*}
\end{center}
\renewcommand{\arraystretch}{1}

\end{document}